\documentclass[%
preprint,
amsmath,amssymb,
aps,pra,showkeys
]{revtex4-2}
\usepackage{graphicx}% Include figure files
\usepackage{dcolumn}% Align table columns on decimal point
\usepackage{bm}% bold math
\usepackage{graphicx,wrapfig}
\usepackage{xcolor}

\DeclareMathOperator{\Tr}{Tr}

\makeatletter
\long\def\footnote#1{%
	\@ifnextchar[\@xfootnote{%
		\stepcounter{footnote}%
		\protected@xdef\@thefnmark{\thefootnote}%
		\@footnotemark\@footnotetext{#1}%
	}%
}
\makeatother

\begin{document}

\title{The polarization behavior of a system of two interacting identical spins 1/2 in adiabatically varying magnetic field}

\author{Yury Belousov}
\email{belousov.yum@phystech.edu}

\author{Dim Safin}
\email{safin.dr@phystech.edu}
%\email{safin.dr@phystech.edu}
\affiliation{%
 Moscow Institute of Physics and Technology \\
	Institutsky lane 9, Dolgoprudny, Moscow region, 141700
}%
\date{\today}

\begin{abstract}

We derive the density matrix of a nonstationary system in the adiabatic representation and obtain the corresponding Liouville equation. This representation is particularly suitable when the standard interaction representation is inapplicable. We illustrate the approach with two identical, spatially separated spin-1/2 particles coupled by a magnetic dipole--dipole interaction in a slowly varying external magnetic field. Specifically, we analyze the spin polarization of pseudo-qutrit states in regimes where the states become entangled. The application of the adiabatic representation to calculate the dependence of observed physical quantities on time for quantum systems is illustrated. Our results show that entangled states can significantly modify the behavior of these observables and provide additional opportunities for their control.

\end{abstract}
\keywords{adiabatic representation; identical spins 1/2; pseudo-qutrit; time-dependent external field; polarization behavior}

\maketitle

\section{Introduction}

The evolution of quantum systems in adiabatically varying external fields is extremely important for many applications in quantum technologies. In particular, adiabatic quantum gates based on the well-known Landau--Zener--Majorana--St\"uckelberg transitions play an important role in quantum information processing (see, for example \cite{A, B, C, MM, AM,PK,GZL}).
Basically, these applications are based on well-known classical solutions of the Schr\"odinger equation, when the Hamiltonian depends on adiabatically varying parameters, \cite{L-ad,Z,M, St,LL} or the Berry phase obtained later \cite{Berry84,Simon,Berry09,ZKP}, observed for a non-degenerate spectrum of a quantum system. As usual, adiabatic approximation is based on the use of time-dependent basis states, when the basis states change in time in accordance with the time change of some parameters of the Hamiltonian, and can be successfully applied in many problems.

Recently \cite{BMG2021,BCM2023}, an `adiabatic representation' was introduced to describe the behavior of coupled spin systems in an adiabatically varying external magnetic field. This representation is applicable for any quantum system with a discrete energy spectrum, including degenerated cases \cite{B2024}. It has been shown that the adiabatic representation makes it possible to find the evolution operator of quantum systems in a varying external magnetic field. This approach is useful for many applications both in quantum technologies and in any other physical problems where a quantum system with a discrete but not necessarily non-degenerate energy spectrum is studied. It was shown in \cite{B2024} that the adiabatic representation is applicable to describe quantum systems consisting of several interacting subsystems, when it is convenient to describe the states of the system in a shared basis. In particular, the obtained evolution operator can be applied to describe the quantum dynamics of two electron spins in a double quantum dot \cite{Hayashi,HuDS,Gorman,Petta,Mason,Giounanlis} and a double quantum well \cite{Anderlini1,Anderlini2}.

For many applications in quantum technologies, it is often important to take into account various relaxation processes when the evolution operator cannot be found. For this reason, it is necessary to construct an adiabatic representation for the density matrix, which is more suitable for calculating physical quantities, in particular, for solving kinetic equations in non-stationary cases.

As shown in \cite{BCM2023,BMG2022}, a system of two interacting identical but separated particles with spin 1/2 can be described as a pseudo-qutrit due to the conservation of total spin. In these works, the evolution of pseudo-qutrit states in the case of an adiabatically time-varying external magnetic field was analyzed in detail. It has been shown that the evolution of states is very different from the standard solution obtained with the adiabatic approach, in cases where the energy levels are close to each other, or in the case of entangled states. Therefore, the behavior of physical quantities, such as spin polarization, should exhibit these features. To determine the spin polarization, we need to find the density matrix of the system. The behavior of entangled states can be clearly described in the adiabatic representation \cite{BMG2021,BCM2023,B2024}. Thus, we need to derive the Liouville equation in an adiabatic representation similar to the Schr\"odinger equation.

This article is organized as follows. In Section \ref{sec2}, the Liouville equation for the density matrix is written in an adiabatic representation and a problem is formulated for two identical interacting 1/2 spins. In Section \ref{sec3}, a solution for the pseudo-qutrit density matrix is obtained and expressions for the longitudinal polarization at two initial states of the system are given.

Lastly, Section \ref{sec4} discusses the application of the adiabatic representation for systems with two interacting spins in an adiabatically varying magnetic field.

\section{Methods}\label{sec2}
\subsection{Effective interaction}\label{subsec2-1}

The adiabatic representation is introduced as follows. In the general case, when the Hamiltonian of the system depends on time, we need to solve the nonstationary  Schr{\"o}dinger equation
\begin{equation}\label{dma1}
   {\rm i}\hbar\frac{\partial}{\partial t}|\psi(t)\rangle = \widehat{H}(t)|\psi(t)\rangle
\end{equation}
with the initial condition $|\psi(t=0)\rangle =|\psi_0\rangle$. If the Hamiltonian depends on time only through some slowly varying in time parameters $\xi(t)$ an adiabatic representation can be introduced.
When adiabatic representation is introduced, one supposes that the stationary Schr{\"o}dinger equation
\begin{equation}\label{dma2}
    \widehat{H}(\xi(t))|\varphi_n(\xi(t))\rangle= E_n(\xi(t))|\varphi_n(\xi(t))\rangle
\end{equation}
can be solved for any moment of time. We assume that the system has a discrete spectrum.

The solution of (\ref{dma2}) gives us the opportunity to use the eigenstates of the Hamiltonian as basis vectors $|\varphi_n(\xi(t))\rangle$ states at any given time. Thus, the solution of (\ref{dma2}) can be written as a superposition at any time:
\begin{equation}\label{dma3}
    |\psi(\xi(t))\rangle=\sum_n a_{n}(\xi(t))|\varphi_n(\xi(t))\rangle, \quad
    {\rm or} \quad |\psi(\xi(t))\rangle =\widehat{\Theta}(\xi(t))|\varphi(\xi(t))\rangle,
\end{equation}
where $\widehat{\Theta}(\xi(t))$ is the unitary operator diagonalizing the Hamiltonian. The unindexed ket $|\varphi(\xi(t))\rangle$ denotes the coefficient vector in this basis.

Using the definition of (\ref{dma3}), we can introduce the adiabatic representation of \cite{BMG2021} to solve the Schr\"odinger equation (\ref{dma1}).
Let us write the Liouville equation for the density  matrix of the nonstationary system in the adiabatic representation. First, we should rewrite the definition of (\ref{dma3}) for the density matrix:
\begin{equation}\label{dma4}
\!\!\! \hat{\rho}_c(t)=|\psi(\xi(t))\rangle\langle\psi(\xi(t))|=
    \widehat{\Theta}(\xi(t))|\varphi(\xi(t))\rangle\langle\varphi(\xi(t))|\widehat{\Theta}^{\dagger}(\xi(t))=
    \widehat{\Theta}(\xi(t))\tilde{\rho}_c(t)\widehat{\Theta}^{\dagger}(\xi(t)),
\end{equation}
where $\tilde{\rho}_c(t)$ is a density matrix in the adiabatic representation. In the definition (\ref{dma4}) the Liouville equation takes a view:
\[
 \dot{\widehat{\Theta}}\tilde{\rho}_c(t)\widehat{\Theta}^{\dagger}+
 \widehat{\Theta}\dot{\tilde{\rho}}_c(t)\widehat{\Theta}^{\dagger}+
  \widehat{\Theta}\tilde{\rho}_c(t){\dot{\widehat{\Theta}}}^{\dagger} +\frac{{\rm i}}{\hbar}
  \left[\widehat{H},\widehat{\Theta}\tilde{\rho}_c(t)\widehat{\Theta}^{\dagger}\right]=0.
\]
Multiplying this equation by $\widehat{\Theta}^{\dagger}$ on the left, we get the following equation:
\begin{equation}\label{dma5}
    \dot{\tilde{\rho}}_c(t) +\frac{{\rm i}}{\hbar}\left[\left(\widehat{\widetilde{H}}-{\rm i}\hbar \widehat{\Theta}^{\dagger}\dot{\widehat{\Theta}}\right),\tilde{\rho}_c \right]=0,
\end{equation}
where the effective unperturbed Hamiltonian $\widehat{\widetilde{H}} =
\widehat{\Theta}^{\dagger}\widehat{H}\widehat{\Theta}$ has a diagonal form for the eigenstates $|\varphi(\xi(t))\rangle$, and the additional operator can be taken into account as a small effective perturbation $\widehat{V}^{eff}=-{\rm i}\hbar\widehat{\Theta}^{\dagger}\dot{\widehat{\Theta}}$. Note, that $\widehat{\Theta}^{\dagger}\widehat{\Theta}=1$ and $\widehat{\Theta}^{\dagger}\dot{\widehat{\Theta}}=-\dot{\widehat{\Theta}}^{\dagger}\widehat{\Theta}$.

We should supplement (\ref{dma5} ) with the initial conditions for the transformed density matrix:
\begin{equation}\label{dma6}
    \tilde{\rho}_c(t)_{\Bigl| t=0}=\widehat{\Theta}^{\dagger}(0)\hat{\rho}_c(0)\widehat{\Theta}(0).
\end{equation}

An average meaning of any physical value in the adiabatic representation has the form:
\begin{equation}\label{dma7}
    \langle \hat{f}\rangle = {\Tr}(\hat{f}\rho_c)= {\Tr}(\hat{f}\widehat{\Theta}\tilde{\rho}_c\widehat{\Theta}^{\dagger})=
    {\Tr}(\widehat{\Theta}^{\dagger}\hat{f}\widehat{\Theta}\tilde{\rho}_c) \equiv
    {\Tr} ({\hat{\tilde f}}\tilde{\rho}_c).
\end{equation}

It is easy to see that the relations \eqref{dma5} -\eqref{dma7} obtained for pure states are valid for the density matrix of an arbitrary ensemble and we can write the basic equation \eqref{dma5} as
\begin{equation*}\label{dma5a}
    \dot{\tilde{\rho}}(t) +\frac{{\rm i}}{\hbar}\left[\left(\widehat{\widetilde{H}}-{\rm i}\hbar \widehat{\Theta}^{\dagger}\dot{\widehat{\Theta}}\right),\tilde{\rho} \right]=0. \eqno{(5}a)
\end{equation*}
Accordingly, Eq.~\eqref{dma7} becomes $\langle\hat f\rangle=\Tr(\widehat{\widetilde f}\,\widetilde\rho)$.

We can solve Eq.~\eqref{dma5a} using the perturbation theory approach. First, it is convenient to use the interaction representation, since the evolution operator has a diagonal form in basis (\ref{dma2}), namely,
\begin{equation}\label{dma8}
\!\!\! \langle \varphi_m(\xi(t))| \widehat{T}\exp\left(-\frac{{\rm i}}{\hbar}\int\limits_{0}^t
    \widehat{\widetilde{H}}(\xi(t')) {\rm d} t'\right)|\varphi_n(\xi(t))\rangle =
    \delta_{mn}\exp\left(-\frac{{\rm i}}{\hbar}\int\limits_{0}^t
    E_n(\xi(t')) {\rm d} t'\right).
\end{equation}
In the interaction representation, all non-diagonal elements of the effective perturbation $\widehat{V}^{eff}$ oscillate with the corresponding transition frequencies:
\begin{equation}\label{dma9}
    \langle \varphi_m(\xi(t))|\widehat{V}^{eff}_{I}|\varphi_n(\xi(t))\rangle =
    V^{eff}_{mn}{\rm e}^{-{\rm i}\int\limits_{0}^t \omega_{mn}(\xi(t')) {\rm d} t'},
\end{equation}
where $\hbar \omega_{mn}(\xi(t)) =E_m(\xi(t))-E_n(\xi(t))$ and $V^{eff}_{mn}=-{\rm i}\hbar\langle \varphi_m(\xi(t))|\widehat{\Theta}^{\dagger}\dot{\widehat{\Theta}}|\varphi_n(\xi(t))\rangle$ are the standard matrix elements, but depending on time.

\subsection{Two interacting identical particles with 1/2 spins }\label{subsec2-2}

Let us consider as an example a system of two identical separated particles with 1/2 spins coupled by a magnetic dipole-dipole interaction. The behavior of this system is equivalent to qutrit, and it has been named pseudo-qutrit \cite{BMG2022}. As was shown in \cite{BCM2023}, the evolution of states reveals very interesting features if the external magnetic field varies adiabatically in the vicinity of values when the states of the two spins are entangled. The best way to observe these features is to study the behavior of the spin polarization of the system. The spin polarization of pseudo-qutrit in a static magnetic field was described in detail in \cite{BMG2022}. Now we will describe the spin polarization of pseudo-qutrit in an adiabatically varying external magnetic field. To begin with, it is useful to recall some main relations.

The Hamiltonian of two identical interacting 1/2 spins in a varying external magnetic field ${\bf B}\|z$ can be written as follows:
\begin{equation}\label{dma10}
 \widehat{H}=-\omega(t)(\hat{s}_{1z} + \hat{s}_{2z}) - 3\Omega ({\mathbf n}\hat{\mathbf s}_1)({\mathbf n}\hat{\mathbf s}_2) +\Omega (\hat{\mathbf s}_1\hat{\mathbf s}_2).
\end{equation}
Here we use units with $\hbar =1$ and $\omega= g\mu_0 B(t), \quad\Omega=(g\mu_0)^2/R^3$, where $g$ and $\mu_0$ are the gyromagnetic ratio and magneton, respectively, $R$ is the distance between two spins, and ${\mathbf n}= {\mathbf R}/R$ is a unit vector. In this system the total spin $S$ is conserved and only triplet states give a contribution to the spin polarization and represent the pseudo-qutrit. The pseudo-qutrit states is suitable to choose as follows
\begin{equation}\label{dma11}
 |q_1\rangle =|+\rangle|+\rangle, \quad
 |q_2\rangle =|-\rangle|-\rangle, \quad
 |q_3\rangle =\frac{1}{\sqrt{2}}(|-\rangle|+\rangle + |+\rangle|-\rangle).
\end{equation}
In the basic states (\ref{dma11}) the Hamiltonian (\ref{dma10}) is determined by the $3\times 3$ matrix
\begin{equation}\label{dma12}
\!\!\! \widehat{H} =
 \frac{\Omega}{4} \left(
  \begin{array}{ccc}
  1-3\cos^2\theta-x(t) & -3\sin^2\theta &
 -\frac{3}{\sqrt{2}}\sin 2\theta \\
  -3\sin^2\theta &1-3\cos^2\theta +x(t) &
  \frac{3}{\sqrt{2}}\sin 2\theta \\
  -\frac{3}{\sqrt{2}}\sin 2\theta & \frac{3}{\sqrt{2}}\sin 2\theta &
  -2(1- 3\cos^2\theta ) \\
  \end{array}
  \right).
\end{equation}
Here $x=4\omega(t)/\Omega$ is a dimensionless external magnetic field.

The Hamiltonian matrix can be approximately diagonalized using the following unitary transformation matrix \cite{BCM2023}:
\begin{equation}\label{dma13}
 \widehat{\Theta} =
  \left(
  \begin{array}{ccc}
 \cos\vartheta_1 & -\sin\vartheta_1\cos\vartheta_2 & \sin\vartheta_1\sin\vartheta_2 \\
  \sin\vartheta_1 & \cos\vartheta_1\cos\vartheta_2 & -\cos\vartheta_1\sin\vartheta_2  \\
  0 & \sin\vartheta_2 & \cos\vartheta_2 \\
  \end{array}
  \right).
\end{equation}
Here parameters of the transformation are determined as follows:
\begin{equation}\label{dma14}
    \tan 2\vartheta_1= \frac{3y}{x}, \quad {\rm where} \quad y = \sin^2\theta,
\end{equation}
and
\begin{equation}\label{dma15}
	\begin{split}
		\tan(2\vartheta_2)={}&
		\frac{3\sqrt{2}\sin(2\theta)S_+}
		{\sqrt{x^2+9y^2}-3(2-3y)},\\
		S_+={}&\cos\vartheta_1+\sin\vartheta_1,\qquad y=\sin^2\theta.
	\end{split}
\end{equation}
The branch is chosen continuously with $\vartheta_2\to0$ as
$x\to+\infty$. For $x>2$ and $0<\theta<\pi/2$, this selects
$0<\vartheta_2<\pi/2$.
When the denominator vanishes while the numerator remains nonzero,
the angle passes continuously through $\vartheta_2=\pi/4$;
the pole of the tangent does not imply a vanishing level splitting.
This rotation eliminates the transformed $2$--$3$ Hamiltonian element
and removes the artificial gap zero produced by the former expression.
It does not generally eliminate the spectator couplings
$\widetilde H_{12}$ and $\widetilde H_{13}$.
Consequently, accuracy of the instantaneous block diagonalization must
be distinguished from accuracy of the resulting polarization dynamics.
The polarization examples below test the analytical evolution within the specified reduced reference model; accuracy against the original Hamiltonian must be assessed separately.

The effective perturbation in (\ref{dma5}) is given by the matrix
\begin{equation}\label{dma16}
 \widehat{\Theta}^{\dagger}\dot{\widehat{\Theta}} =
  \left(
  \begin{array}{ccc}
 0 & -C_2\dot{\vartheta_1} &S_2\dot{\vartheta_1} \\
 C_2\dot{\vartheta_1} & 0 & -\dot{\vartheta}_2 \\
 -S_2\dot{\vartheta_1} & \dot{\vartheta}_2 & 0 \\
  \end{array}
  \right).
\end{equation}
For the sake of brevity, here and further we have introduced the notation
\begin{equation}\label{dma17}
    C_1 =\cos\vartheta_1, \quad S_1 =\sin\vartheta_1, \quad C_2 =\cos\vartheta_2, \quad S_2 =\sin\vartheta_2.
\end{equation}

Further, we will be interested only in the longitudinal component of spin polarization, since it reveals more clearly all the features of the system's behavior in varying magnetic fields. To do this, we will calculate the matrix of the $z$-projection of the full spin in the adiabatic representation:
\begin{align}\label{dma18}
\widehat{\widetilde S}_z(t) =&  \widehat{\Theta}^{\dagger}(t)
    \left(
  \begin{array}{ccc}
 1 &0&0 \\
 0&-1&0 \\
 0&0& 0 \\
  \end{array}
  \right)
    \widehat{\Theta}(t) \nonumber \\
 =&    \left(
  \begin{array}{ccc}
 \cos 2\vartheta_{t,1} &-\sin2\vartheta_{t,1}\cos\vartheta_{t,2} &\sin2\vartheta_{t,1}\sin\vartheta_{t,2} \\
 -\sin2\vartheta_{t,1}\cos\vartheta_{t,2} &-\frac{1}{2}\cos 2\vartheta_{t,1}(1+\cos 2\vartheta_{t,2})
 & \frac{1}{2}\cos 2\vartheta_{t,1} \sin 2\vartheta_{t,2} \\
 \sin2\vartheta_{t,1}\sin\vartheta_{t,2} &\frac{1}{2}\cos 2\vartheta_{t,1} \sin 2\vartheta_{t,2}
 & -\frac{1}{2}\cos 2\vartheta_{t,1}(1-\cos 2\vartheta_{t,2}) \\
  \end{array}
  \right).
\end{align}
Here and further, we denote for brevity $\vartheta_{1,2}(t)\equiv \vartheta_{t,1,2}$.

\section{Results}\label{sec3}
\subsection{Density matrix solution}

Now we can find a solution for the density matrix $\tilde{\rho}_i(t)=U^{\dagger}(t)\tilde{\rho}_c(t)U(t)$ in the standard interaction representation using the solutions obtained in Sections \ref{subsec2-1} and \ref{subsec2-2}:
\begin{equation}\label{dma19}
    \dot{\tilde{\rho}}_I(t) = -{\rm i}\left[\widehat{V}^{eff}_I(t), \tilde{\rho}_I\right].
\end{equation}
First, we will neglect relaxation processes in order to obtain the main features of the evolution of the density matrix in the adiabatic representation. Equation (\ref{dma19}) is equivalent to a system of five linear coupled equations, which is conveniently written in the basis of (\ref{dma11}). After some simple but rather cumbersome calculations, we obtain the following system of equations for the components of the density matrix:
\begin{align}\label{dma20}
 \dot{\rho}_{11} =& \,\dot{\vartheta}_1\left[ C_2\left(\rho_{12}{\rm e}^{{\rm i}\phi_{12}(t)} + h.c.\right)-  S_2\left(\rho_{13}{\rm e}^{{\rm i}\phi_{13}(t)} + h.c.\right)\right], \nonumber \\
    \dot{\rho}_{22} =& -\dot{\vartheta}_1 C_2\left(\rho_{12}{\rm e}^{{\rm i}\phi_{12}(t)} + h.c.\right)+
    \dot{\vartheta}_2\left(\rho_{23}{\rm e}^{{\rm i}\phi_{23}(t)} + h.c.\right),  \nonumber \\
    \dot{\rho}_{33} =& \,\dot{\vartheta}_1S_2\left(\rho_{13}{\rm e}^{{\rm i}\phi_{13}(t)} + h.c.\right)
    -\dot{\vartheta}_2\left(\rho_{23}{\rm e}^{{\rm i}\phi_{23}(t)} + h.c.\right), \\
    \dot{\rho}_{12}+\dot{\rho}_{21} =& -\dot{\vartheta}_1\left[2C_2\cos\phi_{12}(t)\left(\rho_{11}-\rho_{22}\right)+
    S_2\left(\rho_{23}{\rm e}^{{\rm i}\phi_{23}(t)} + h.c.\right)\right] +\dot{\vartheta}_2\left(\rho_{13}{\rm e}^{{\rm i}\phi_{13}(t)} + h.c.\right), \nonumber \\
    \dot{\rho}_{13}+\dot{\rho}_{31} =& \, \dot{\vartheta}_1\left[2S_2\cos\phi_{13}(t)\left(\rho_{11}-\rho_{33}\right)+
    C_2\left(\rho_{23}{\rm e}^{{\rm i}\phi_{23}(t)} + h.c.\right)\right] -\dot{\vartheta}_2\left(\rho_{12}{\rm e}^{{\rm i}\phi_{12}(t)} + h.c.\right), \nonumber \\
    \dot{\rho}_{23}+\dot{\rho}_{32} =& -\dot{\vartheta}_1\left[
    C_2\left(\rho_{13}{\rm e}^{{\rm i}\phi_{13}(t)} + h.c.\right) -S_2\left(\rho_{12}{\rm e}^{{\rm i}\phi_{12}(t)} + h.c.\right)\right] -2\dot{\vartheta}_2\cos\phi_{23}(t)\left(\rho_{22}-\rho_{33}\right).\nonumber
\end{align}
Here we have simplified some notations for brevity. Namely, all the components of the density matrix are represented in the interaction representation: $\rho_{mn}\equiv \tilde{\rho}_{I,mn}$ and
\[
  \phi_{mn}(t) = \int\limits_0^t\omega_{mn}(t') {\rm d} t'.
\]
It is easy to see that $\dot{\rho}_{11}+\dot{\rho}_{22}+\dot{\rho}_{33}=0$ and we have 5 independent equations, but now it is advisable to write this system as a system of 6 equations (\ref{dma20}).

Let us introduce one convenient notation for the non-diagonal elements of the density matrix
\begin{equation}\label{dma21}
  \tilde{\rho}_{ik}=\tilde{\rho}^{*}_{ki}=\tilde{\rho}_{ik}^s(t){\rm e}^{{\rm i}\varphi_{ik}},
\end{equation}
where $\tilde{\rho}_{ik}^s(t)$ is the real amplitude, which varies slowly over time.
In these notations we can write
\[
  \rho_{ik}{\rm e}^{{\rm i}\phi_{ik}(t)} + h.c. = 2\tilde{\rho}_{ik}^s(t)\cos(\phi_{ik}(t)+\varphi_{ik}).
\]
Now, we can rewrite the system of equations \eqref{dma20} as follows
\begin{align}\label{dma22}
 \dot{\tilde{\rho}}_{11} =& \, 2\dot{\vartheta}_1\left( C_2\tilde{\rho}_{12}^s\cos(\phi_{12}(t)+\varphi_{12}) -
    S_2\tilde{\rho}_{13}^s\cos(\phi_{13}(t)+ \varphi_{13})\right), \nonumber \\
    \dot{\tilde\rho}_{22} =& -2\dot{\vartheta}_1 C_2\tilde{\rho}_{12}^s\cos(\phi_{12}(t)+\varphi_{12}) + 2\dot{\vartheta}_2\tilde{\rho}_{23}^s\cos(\phi_{23}(t)+\varphi_{23}), \nonumber \\
    \dot{\tilde\rho}_{33} =& \, 2\dot{\vartheta}_1 S_2\tilde{\rho}_{13}^s\cos(\phi_{13}(t)+\varphi_{13})
    -2\dot{\vartheta}_2\tilde{\rho}_{23}^s\cos(\phi_{23}(t)+\varphi_{23}),  \\
    \dot{\tilde{\rho}}_{12}^s\cos\varphi_{12} =&\!\! -\dot{\vartheta}_1\left[C_2\cos\phi_{12}(t)\left(\rho_{11}\!-\!\rho_{22}\right)\!+\!
    S_2\tilde{\rho}_{23}^s\cos(\phi_{23}(t)\!+\!\varphi_{12}) \right] \! +\dot{\vartheta}_2\tilde{\rho}_{13}^s\cos(\phi_{13}(t)\!+\!\varphi_{13}), \nonumber \\
    \dot{\tilde{\rho}}_{13}^s\cos\varphi_{13} =&\!\, \dot{\vartheta}_1\left[S_2\cos\phi_{13}(t)\left(\rho_{11}\!-\!\rho_{33}\right)\!+\!
    C_2\tilde{\rho}_{23}^s\cos(\phi_{23}(t)\!+\!\varphi_{23}) \right]\! -\!\dot{\vartheta}_2\tilde{\rho}_{12}^s\cos(\phi_{12}(t)\!+\!\varphi_{12}), \nonumber \\
    \dot{\tilde{\rho}}_{23}^s\cos\varphi_{32} =&\! -\dot{\vartheta}_1\left[
    C_2\tilde{\rho}_{13}^s\cos(\phi_{13}(t)\!+\!\varphi_{13})\! -\! S_2\tilde{\rho}_{12}^s\cos(\phi_{12}(t)\!+\!\varphi_{12}) \right]\! -\dot{\vartheta}_2\cos\phi_{23}(t)\left(\rho_{22}\!-\!\rho_{33}\right).\nonumber
\end{align}

It is easy to see that $\dot{\tilde\rho}_{ik}\sim \dot{\vartheta}_{1,2}$ and $\omega_{ik}\gg \dot{\vartheta}_{1,2}$, except of the energy levels are close to each other.
If the inequality $\omega_{ik}\gg\dot{\vartheta}_{1,2}$ is valid, we can integrate the equations of the system (\ref{dma22}), considering the integral functions to be constant and equal to their values at the current time, with the exception of a rapidly oscillating multiplier. For example,
\begin{align}\label{dma23}
  \tilde{\rho}_{11}(t)=&\tilde{\rho}_{11}(0)+2\dot{\vartheta}_1(t)\left( C_2(t)\tilde{\rho}_{12}^s(t)
    \int\limits_0^t\cos(\phi_{12}(t')+\varphi_{12}){\rm d}t' - \right. \nonumber \\
    &\left. S_2(t)\tilde{\rho}_{13}^s(t)\int\limits_0^t\cos(\phi_{13}(t')+ \varphi_{13}){\rm d}t'\right).
\end{align}
Note that
\[
\!\!\! \int\limits_0^t\cos(\phi_{ik}(t')+\varphi_{ik}){\rm d}t'=
  \int\limits_0^{\phi(t)}\cos(\phi_{ik}+\varphi_{ik})\frac{{\rm d}\phi_{ik}}{{\rm d}\phi_{ik}/dt'}=
  \int\limits_0^{\phi(t)}\frac{\cos(\phi_{ik}+\varphi_{ik})}{\omega_{ik}}{\rm d}\phi_{ik}
\]
and due to the small parameter $\dot{\vartheta}_{1,2}/\omega_{ik}\ll 1$, all time-dependent terms on the right side are negligible, and we get
\[
  \tilde{\rho}_{11}(t)\approx\tilde{\rho}_{11}(0),
\]
which strictly corresponds to the theory of adiabatic perturbations for a non-degenerate energy spectrum.

\begin{figure}
\centering
\includegraphics[height=7cm]{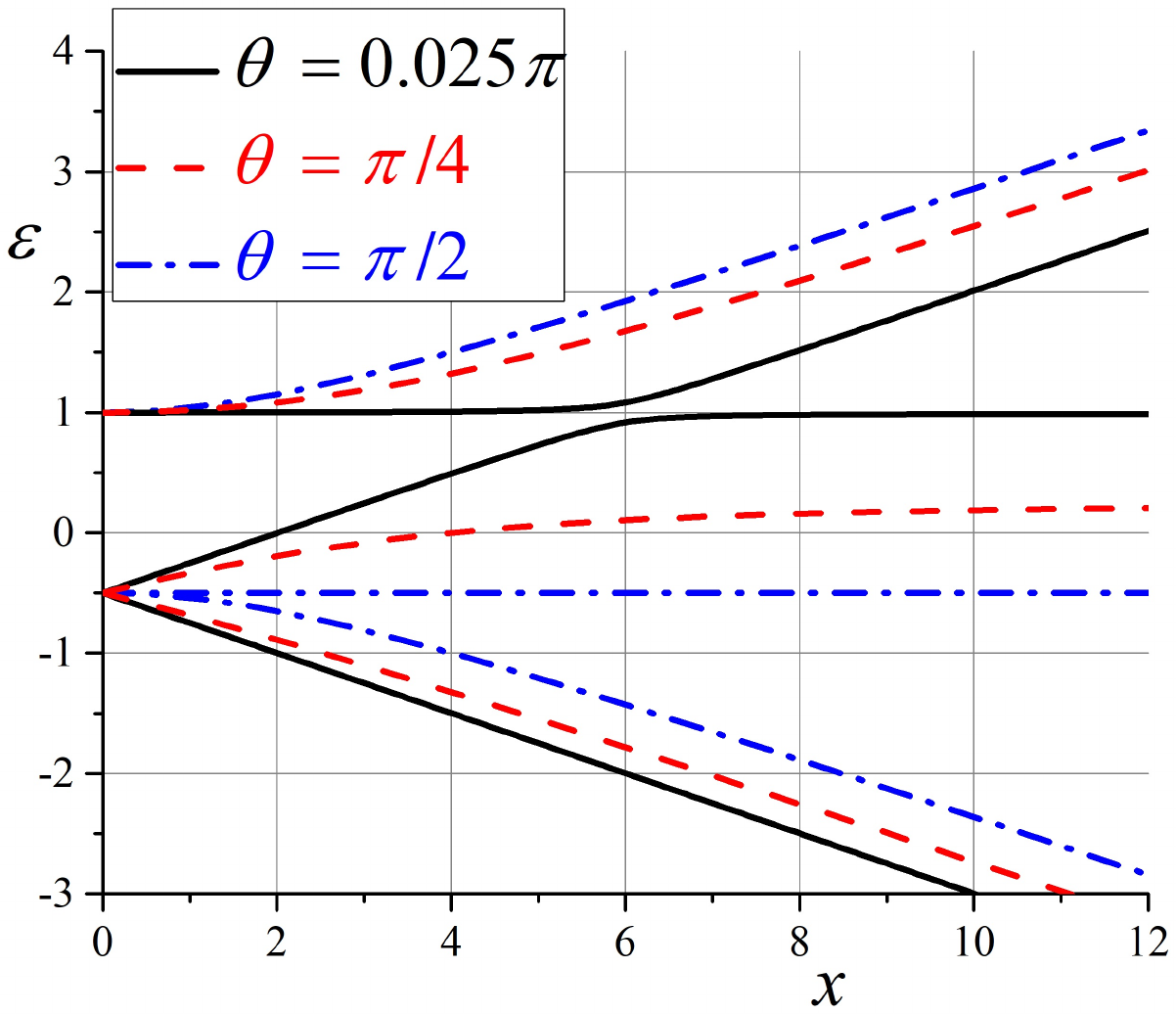}
\caption{Scheme of energy levels. Dependence on the external magnetic field for three different angles $\theta=0.025\pi, \pi/4$, and $\pi/2$.} \label{fig:1}
\end{figure}

But the situation changes a lot if these two levels are close to each other or intercross. The structure of energy levels shows (see figure \ref{fig:1}, \cite{BCM2023}) that this picture occurs in small external magnetic fields for levels $\varepsilon_1$ and $\varepsilon_2$ and in intermediate fields for levels $\varepsilon_2$ and $\varepsilon_3$. In these cases, our system of equations \eqref{dma22} reduces to systems of two equations.
%%%%%%%%%%%%%%%%%%%%%%%%%%%%%%%%%%%%%%%%%%%%%%%%

%%%%%%%%%%%%%%%%%%%%%%%%%%%%%%%%%%%%%%%%%%%%%%%%%%%%%%%%%%%%%%

Indeed, in the case of $x\ll 1$, we have
\begin{align}\label{dma24}
  \frac{{\rm d}}{{\rm d} t}(\tilde{\rho}_{11}-\tilde{\rho}_{22}) =&  4\dot{\vartheta}_1
  C_2\tilde{\rho}_{12}^s\cos(\phi_{12}(t)+\varphi_{12}), \nonumber \\
  \frac{{\rm d}}{{\rm d} t}\tilde{\rho}_{12}^s = & -\frac{\dot{\vartheta}_1C_2}{\cos\varphi_{12}}\cos\phi_{12}(t)\left(\tilde{\rho}_{11}- \tilde{\rho}_{22}\right)
\end{align}
The equation for $\tilde{\rho}_{33}$ is not independent, and solutions for the components $\tilde{\rho}_{13}^s$ and $\tilde{\rho}_{23}^s$ can be obtained when the solution of the system (\ref{dma24}) is known. Namely,
\begin{align}\label{dma25}
  \tilde{\rho}_{33}(t)= & \,1-(\tilde{\rho}_{11}(t) + \tilde{\rho}_{22}(t)) \approx \tilde{\rho}_{33}(0), \nonumber \\
  \tilde{\rho}_{11}(t)=& \, \tilde{\rho}_{11}(0)+ 2\int\limits_0^t \dot{\vartheta}_1 C_2\tilde{\rho}_{12}^s\cos(\phi_{12}(t')+\varphi_{12}){\rm d}t', \nonumber \\
  \tilde{\rho}_{13}(t)^s =& \, \tilde{\rho}_{13}(0) - \int\limits_0^t \frac{\dot{\vartheta}_2}{\cos\varphi_{13}} \cos(\phi_{12}(t')+\varphi_{12})\tilde{\rho}_{12}^s {\rm d} t', \nonumber \\
  \tilde{\rho}_{23}^s(t) =& \, \tilde{\rho}_{23}(0) + \int\limits_0^t \frac{\dot{\vartheta}_1S_2}{\cos\varphi_{12}} \cos(\phi_{12}(t')+\varphi_{12})\tilde{\rho}_{12}^s {\rm d} t'.
\end{align}

In the case of an intermediate external field, the components $\tilde{\rho}_{22}$ and $\tilde{\rho}_{33}$ are interconnected, and we have the following system of equations:
\begin{align}\label{dma26}
  \frac{{\rm d}}{{\rm d} t}(\tilde{\rho}_{22}-\tilde{\rho}_{33}) = \, 4\dot{\vartheta}_2
  \tilde{\rho}_{23}^s\cos(\phi_{23}(t)+\varphi_{23}), \nonumber \\
  \frac{{\rm d}}{{\rm d} t}\tilde{\rho}_{23}^s = -\frac{\dot{\vartheta}_2}{\cos\varphi_{23}}\cos\phi_{23}(t)\left(\tilde{\rho}_{22}- \tilde{\rho}_{33}\right).
\end{align}
Diagonal component of $\tilde{\rho}_{11}(t) \approx \tilde{\rho}_{11}(0)$, and the other non-diagonal components are equal to:
\begin{align}
    \tilde{\rho}_{12}^s(t) =& \,\tilde{\rho}_{12}(0) - \int\limits_0^t \frac{\dot{\vartheta}_1S_2}{\cos\varphi_{12}} \cos(\phi_{23}(t')+\varphi_{23})\tilde{\rho}_{23}^s {\rm d} t', \nonumber \\
  \tilde{\rho}_{13}^s(t) =& \,\tilde{\rho}_{13}(0)+ \int\limits_0^t \frac{\dot{\vartheta}_1C_2}{\cos\varphi_{13}} \cos(\phi_{23}(t')+\varphi_{23})\tilde{\rho}_{23}^s {\rm d}t', \nonumber \\
  \tilde{\rho}_{22}(t) =& \,\tilde{\rho}_{22}(0) + 2\int\limits_0^t \dot{\vartheta}_2\tilde{\rho}_{23}^s\cos(\phi_{23}(t')+\varphi_{23}) {\rm d} t'.\label{dma27}
\end{align}
We can choose the phases $\varphi_{ik}=0$ in the notation \eqref{dma21}, and the equations \eqref{dma24} and \eqref{dma26} can be written in universal form if we introduce a two-dimensional  vector ${\vec \rho}$ with components ${\vec\rho}\,^{T}= (\tilde{\rho}_{ii}-\tilde{\rho}_{kk}, 2\tilde{\rho}_{ik}^s )$:
\begin{equation}\label{dma28}
    \frac{d}{dt}{\vec\rho}=i2{\dot \vartheta}\cos\phi_{ik}(t)\sigma_y {\vec\rho},
\end{equation}
where $\sigma_y$ is the Pauli operator, and ${\dot\vartheta}= \dot{\vartheta}_1 \cos\vartheta_2$ for \eqref{dma24} and ${\dot\vartheta}= \dot{\vartheta}_2$ for \eqref{dma26}

\subsection{Consideration of relaxation processes}

Relaxation processes can strongly change the behavior of pseudo-qutrit polarization. Therefore, it is useful to take into account the relaxation in the Liouville equation for the density matrix. To do this correctly, we need to use the Master Equations which for a system with discrete and non-degenerated spectrum have the form (see, e.g. \cite{Blum,Fine})
\begin{equation}\label{dma29}
	\begin{split}
		\dot\rho_{I,mn}+
		\frac{i}{\hbar}[\Gamma,\rho_I]_{mn}
		={}&\delta_{mn}\sum_{k\ne m}
		W_{mk}\rho_{I,kk} - \gamma_{mn}\rho_{I,mn}.
	\end{split}
\end{equation}
Here $\rho_I$ is the density matrix in the interaction representation. All matrices are determined by interactions with the environment, but here we will use their matrix elements as some phenomenological parameters, since we are trying to trace some general features of the influence of relaxation processes on the behavior of polarization. This phenomenological, weak-coupling (Markovian) treatment is the standard one for $W$, $\gamma \ll \omega_{ik}$, and its range of applicability has recently been mapped experimentally: in a tunable superconducting flux qubit swept through an anticrossing, the dominant open-system effect was shown to cross over from a weakly coupled Markovian environment to a strongly coupled non-Markovian one as
the tunneling amplitude is reduced \cite{Dai2024}. In that work a Lindblad-type adiabatic master equation of the form \eqref{dma29} reproduces the data on the weak-coupling side, while it fails once the time the system needs to traverse the anticrossing becomes comparable with the reorganization time of the low-frequency environment. In the present problem we stay well inside the first regime, since $W, \gamma \lesssim 0.2 \Omega$ while $\omega_{ik} \sim \Omega$. In the future, we will neglect the shifts in energy levels determined by the matrix elements $\Gamma$. Note that $\gamma_{mn}=\gamma_{nm}$ and $W_{mk} \neq W_{km}$, and $W_{mk}/W_{km}=\exp((E_k-E_m)/T)$. When the couplings to the remaining state can be neglected at the retained order, the dynamics of a selected pair can be treated within a reduced set of equations. This reduction requires a separation-of-scales argument; it does not follow from entanglement alone. We add the phenomenological pair-relaxation terms to Eqs.~\eqref{dma24} and \eqref{dma26} as follows:
\begin{align}\label{dma30}
  \frac{{\rm d}}{{\rm d} t}(\tilde{\rho}_{ii}-\tilde{\rho}_{kk}) =& 4\dot{\vartheta}
  \tilde{\rho}_{ik}^s\cos\phi_{ik}(t)- 2 W_{ki}\tilde{\rho}_{ii}
  +2 W_{ik}\tilde{\rho}_{kk}, \nonumber \\
  \frac{{\rm d}}{{\rm d} t}\tilde{\rho}_{ik}^s = & -\dot{\vartheta}\cos\phi_{ik}(t)\left(\tilde{\rho}_{ii}- \tilde{\rho}_{kk}\right) - \gamma_{ik} \tilde{\rho}_{ik}^s.
\end{align}

For an isolated active pair without population transfer to a spectator,
\begin{equation*}
	\dot\rho_{ii}+\dot\rho_{kk}=0,\qquad
	\rho_{ii}(t)+\rho_{kk}(t)=\rho_{ii}(0)+\rho_{kk}(0).
\end{equation*}

For the two-dimensional vector ${\vec\rho}$ introduced above, equation (\ref{dma30}) can be written as:
\begin{equation}\label{dma31}
    \frac{d}{dt}{\vec\rho}=\left[i2{\dot \vartheta}\cos\phi_{ik}(t)\sigma_y - \Gamma\sigma_z -\Gamma_0 \right]{\vec\rho},
\end{equation}
where $\Gamma = (W-\gamma)/2,\quad \Gamma_0=(W+\gamma)/2,\quad W= W_{ik}
+ W_{ki}$.

The solution of the system of equations (\ref{dma31}) in the case of neglect of relaxation processes when $\Gamma$ and $\Gamma_0$ are zero was obtained in the previous section. If we denote this solution as ${\vec\rho}_0(t)$, then we can write
\begin{equation}\label{dma32}
 {\vec\rho}_0(t) = {\widehat G}(t,0) {\vec\rho}(0), \quad \mbox{where} \quad
   {\widehat G}(t,0) = \exp\left[i2\int\limits_0^t {\dot \vartheta}(t')\cos\phi_{ik}(t')dt'\sigma_y\right].
\end{equation}
Using this expression, we can represent the solution of the system (\ref{dma31}) in a compact form if we introduce the ``interaction'' representation
\begin{equation}\label{dma33}
    {\vec\rho}(t) = {\widehat G}(t,0){\vec\rho}^G(t).
\end{equation}
Indeed, the operator ${\widehat G}(t,0)$ is unitary: ${\widehat G}^{-1}(t,0)={\widehat G}^{\dagger}(t,0)$, hence Eq. (\ref{dma32}) for the vector ${\vec\rho}^G(t)$ is transformed to the following form:
\begin{equation}\label{dma34}
 \frac{d}{dt}{\vec\rho}^G=-\left(\Gamma_0+{\widehat\Gamma}_G(t)\right){\vec\rho}^G, \quad
     {\widehat\Gamma}_G(t)=\Gamma {\widehat G}^{\dagger}(t,0)\sigma_z {\widehat G}(t,0), \quad
     {\vec\rho}^G(0)={\vec\rho}(0).
\end{equation}
Thus, the operator ${\widehat G}(t,0)$ can be interpreted as an intermediate evolution operator, and the operator ${\widehat\Gamma}_G(t)$ is equal to
\begin{equation}\label{dma35}
	\widehat\Gamma_G(t)=\Gamma\left[
	\cos(2\Phi_{ik}(t))\sigma_z+
	\sin(2\Phi_{ik}(t))\sigma_x\right],
\end{equation}
where $\Phi_{ik}(t)=\int_0^t\dot\vartheta(t')
\cos\phi_{ik}(t')\,dt'$.

The exact time-ordered representation of Eq.~\eqref{dma34} is
\begin{equation}\label{dma36}
	\begin{split}
		\vec\rho^G(t)&=e^{-\Gamma_0t}\widehat U_R(t,0)\vec\rho(0),\\
		\widehat U_R(t,0)&=\widehat T\exp\left\{
		-\Gamma\int_0^t\left[
		\cos(2\Phi_{ik}(t'))\sigma_z+
		\sin(2\Phi_{ik}(t'))\sigma_x\right]dt'\right\}.
	\end{split}
\end{equation}
In general, the operator ${\widehat U}_{R}$ for a two-level system can be represented as
\begin{equation}\label{dma37}
    {\widehat U}_{R} = 1-\Gamma\left(f_0(t)+f_z(t)\sigma_z +f_x(t)\sigma_x + if_y(t)\sigma_y\right),
\end{equation}
and we need to find four unknown functions $f_0(t),\,f_z(t),\,f_x(t),\,f_y(t)$.

The specific features of the polarization behavior can be effectively observed in the case of relatively slow relaxation compared to oscillation frequencies, namely at $\Gamma\ll\omega_{ik}$. In this case, we can get a good approximate expression for the vector ${\vec\rho}_G(t)$ if we limit ourselves to only the first two orders of decomposition of ${\widehat U}_{R}$. For the sake of brevity, we introduce the following notation:
\begin{equation}\label{dma38}
    \lambda_1 = \frac{\dot \vartheta}{\omega_{ik}} \ll 1 \quad \mbox{and}\quad
    \lambda_2 = \frac{\Gamma}{\omega_{ik}} \ll 1.
\end{equation}

Using the functions (\ref{a2}) given in Appendix 1, we get for ${\vec \rho}^R(t)={\widehat U}_{R}{\vec\rho}(0)$:
\begin{align}
 (\rho_{ii}(t)-\rho_{kk}(t))^R =& \left[1-\Gamma\left(1-\lambda_1^2\right)t+\frac{1}{2}\Gamma^2\left(1-\lambda_1^2\right)^2t^2 \right](\rho_{ii}(0)-\rho_{kk}(0)) -\nonumber \\
  -& 4\left(1-\lambda_2\right)\lambda_1\lambda_2 \rho_{ik}(0) \nonumber \\
  \rho_{ik}^R(t) =&\left[1+\Gamma\left(1-\lambda_1^2\right)t+\frac{1}{2}\Gamma^2\left(1-\lambda_1^2\right)^2t^2 \right]\rho_{ik}(0) + \nonumber \\
  +& \Gamma\left(1-\lambda_2\right)\lambda_1\lambda_2  (\rho_{ii}(0)-\rho_{kk}(0)).\label{dma39}
\end{align}

It can be see that the expression in square brackets is an expansion of an exponential with second-order precision in a small parameter. Thus, we get the solution of Eq.\eqref{dma36} in the form:
\begin{align}
\left(\rho_{ii}(t)-\rho_{kk}(t)\right)^G =& e^{-\left(W -\Gamma\lambda_1^2\right)t}(\rho_{ii}(0)-\rho_{kk}(0)) -   2e^{-\Gamma_0t}\left(1-\lambda_2\right)2\lambda_1\lambda_2 \rho_{ik}(0) \nonumber\\
  \rho^G_{ik}(t) =& e^{-\left(\gamma -\Gamma\lambda_1^2 \right)t}\rho_{ik}(0)+e^{-\Gamma_0t} \left(1-\lambda_2\right) 2\lambda_1\lambda_2 (\rho_{ii}(0)-\rho_{kk}(0)).
\end{align}

Finally, the solution of the system \eqref{dma31} can be written as \footnote{The following implemented second-order expressions must be interpreted
	within the small-parameter regime and checked against numerical
	integration of the same reduced equations.
	The parameters $\lambda_1$ and $\lambda_2$ are evaluated without clipping;
	values of order unity indicate loss of perturbative control rather
	than a numerical instability to be regularized.
	In addition, the phase convention and initial-time boundary terms of
	the derivation require a consistent check before these expressions
	can be identified with a revised expansion of Eq.~\eqref{dma36}.}
\begin{align}
& \rho_{ii}(t)-\rho_{kk}(t) = \nonumber \\
 & e^{-Wt}\left[e^{\Gamma\lambda_1^2 t} \cos 2\Phi_{ik}(t) + e^{-\Gamma t}\left(1-\lambda_2\right) 2\lambda_1\lambda_2\sin 2\Phi_{ik}(t) \right] (\rho_{ii}(0)-\rho_{kk}(0)) -\nonumber\\
 -&e^{-\gamma t}\left[e^{\Gamma\lambda_1^2 t}\sin 2\Phi_{ik}(t) -  e^{-\Gamma t}\left(1-\lambda_2\right)\lambda_1\lambda_2\cos 2\Phi_{ik}(t)\right] \rho_{ik}(0),
 \label{dma41} \\
 & \rho^s_{ik}(t) =  \nonumber\\
   & e^{-\gamma t}\left[ e^{\Gamma\lambda_1^2 t}
  \cos 2\Phi_{ik}(t) +
  e^{-\Gamma t}\left(1-\lambda_2\right) 2\lambda_1\lambda_2\sin2\Phi_{ik}(t) \right]\rho_{ik}(0)+
  \nonumber\\
  +& e^{-\Gamma_0t}\left[\left(1-\lambda_2\right) 2\lambda_1\lambda_2\cos 2\Phi_{ik}(t) -
  e^{-\Gamma\left(1 -\lambda_1^2 \right)t} \sin 2\Phi_{ik}(t) \right] (\rho_{ii}(0)-\rho_{kk}(0)).\label{dma42}
\end{align}

\subsection{Longitudinal polarization of pseudo-qutrit}

The spin polarization of the qutrit is determined by the well-known relation $P_i(t)= {\Tr} \widehat{S}_i \hat{\rho}_c(t) = {\Tr} \widetilde{\widehat{S}}_i U_0(t)\tilde{\hat{\rho}}_I(t)U_0^{\dagger}(t)$, where $\widehat{S}_i$ is an $i$-projection of the total spin operator. The evolution operator $U_0(t)$ of an unperturbed Hamiltonian is defined by the relation \eqref{dma8}. We will be interested in the $z$-projection only, since the effects of transitions are more clearly revealed in the behavior of longitudinal polarization. The expression depends on both the initial polarization and the magnitude of the external field, as follows from \eqref{dma23} -- \eqref{dma26}.

Let the initial state of pseudo-qutrit be determined by a diagonal density matrix with elements $\rho_{kk}(0)=w_k$, which defines some probability distribution. The transformed density matrix at the initial moment of time is determined by the following $3\times 3$ matrix
\begin{align}\label{dma44}
    & \tilde{\rho}(0)= \Theta^{\dagger}(0)\rho(0)\Theta(0)\\
    =& \left(
         \begin{array}{ccc}
           w_1C_1^2+w_2S_1^2 & -(w_1-w_2)C_1S_1C_2 & (w_1-w_2)C_1S_1S_2 \\
           -(w_1-w_2)C_1S_1C_2 & (w_1S_1^2+w_2C_1^2)C_2^2+w_3S_2^2 & -[(w_1S_1^2+w_2C_1^2)-w_3]C_2S_2 \\
           (w_1-w_2)C_1S_1S_2 &  -[(w_1S_1^2+w_2C_1^2)-w_3]C_2S_2  & (w_1S_1^2+w_2C_1^2)S_2^2+w_3C_2^2 \\
         \end{array}
       \right) \nonumber
\end{align}
We can see that all non-diagonal components of the density matrix are real and all the phases that should be determined from the initial conditions $\varphi_{ik}=0$. These conditions simplify the relations \eqref{dma23} -- \eqref{dma26}.

Now we can write the longitudinal polarization of the pseudo-qutrit in the general case:
\begin{align}\label{dma45}
 &P_z(t)= \cos 2\vartheta_{t,1}\left[\tilde{\rho}_{11}- \frac{1}{2}(\tilde{\rho}_{22}+\tilde{\rho}_{33})- \frac{1}{2}\cos 2\vartheta_{t,2}(\tilde{\rho}_{22}-\tilde{\rho}_{33}) \right] - \\
 -&2\sin 2\vartheta_{t,1}(\cos \vartheta_{t,2}\tilde{\rho}^{s}_{12}\cos\phi_{12}(t) -
 \sin \vartheta_{t,2}\tilde{\rho}^{s}_{13}\cos\phi_{13}(t)) + \cos 2\vartheta_{t,1}
 \sin 2\vartheta_{t,2}\tilde{\rho}^{s}_{23}\cos\phi_{23}(t).\nonumber
\end{align}

In the following, we will consider the behavior of longitudinal polarization in the intermediate case for $1< x \lesssim 10$, when the density matrix is described by Eq. \eqref{dma26}. As shown, in this case $\tilde{\rho}_{11}(t) \approx \tilde{\rho}_{11}(0)$ and if we take into account the relaxation processes
\begin{equation}\label{dma46}
	\tilde{\rho}_{11}(t) \approx \tilde{\rho}^T_{11}\left(1-e^{-W_1t}\right) + \tilde{\rho}_{11}(0)e^{-W_1t}.
\end{equation}
Here we can suppose in accordance with Eq.\eqref{dma29} that $W_1 =W_{12}+W_{13}$. Respectively,
\begin{equation}\label{dma47}
	\tilde{\rho}_{22}(t)+\tilde{\rho}_{33}(t) \approx (\tilde{\rho}^T_{22}+\tilde{\rho}^T_{33}) \left(1-e^{-Wt}\right) + (\tilde{\rho}_{22}(0)+\tilde{\rho}_{33}(0))e^{-Wt},
\end{equation}
where $\tilde{\rho}^T_{ii}$ the components of  the density matrix in thermal equilibrium.

When source terms for the spectator coherences are neglected, the reduced prescription assigns their signed amplitudes the form
\begin{equation}\label{dma48}
    \tilde{\rho}^s_{12}(t) = \tilde{\rho}^s_{12}(0)e^{-\gamma_1 t}, \quad
    \tilde{\rho}^s_{13}(t) = \tilde{\rho}^s_{13}(0)e^{-\gamma_1 t}.
\end{equation}
Here we assume that this components decay at the same phenomenological rate $\gamma_1$.

Expressions for all the components of the transformed density matrix at the initial time are given in Appendix 3.

\section{Discussions}\label{sec4}

The obtained equations give us the opportunity to examine the behavior of the  spin polarization of pseudo-qutrit in varying external magnetic fields and determine its dependence for the cases of separable or entangled states of two spins.

For numerical calculations, we used a model of an adiabatically varying external magnetic field in the form
\begin{equation}\label{two39}
	x(\tau)=x_0+a\tanh[\alpha(\tau-\tau_0)],
\end{equation}
Here $\tau=\Omega t$. We use $x_0=6$, $a=4$, $\alpha=0.1$, and $\tau_0=100$, and follow the evolution over $0\leq\tau\leq200$. The field-change rate is maximal at $\tau=\tau_0$, where $x=6$. The endpoint fields are $x(0)=6-4\tanh 10$ and $x(200)=6+4\tanh 10$, approximately 2 and 10, respectively. The central and late windows are $90\leq\tau\leq110$ and $140\leq\tau\leq160$. The same trajectories are used in the complete view and both enlargements.

For an adiabatically varying magnetic field, one needs to choose $\alpha \ll 1$, when the model coincides with the Landau--Majorana--Stuckelberg--Zener (LMSZ) model \cite{L-ad,Z,St,M} widely used in various problems with a time-dependent Hamiltonian \cite{C,VIV07,NSMM23,I,Gr,GVSM23}. Dissipative LMSZ passages of this kind have been realized and characterized experimentally in superconducting circuits, where both the sweep rate and the system-environment coupling can be tuned independently \cite{Dai2024}.

Before proceeding with calculations for arbitrary cases, we tested our analytical results in the case of $\theta=\pi/2$, when the exact energy spectrum at the current time is known. We have obtained an absolutely exact match between the analytical results and the numerical calculations, so we do not present them as a special figure. In addition, we compared the results of the numerical solution for the complete 3-level system and the reduced analytical solution and found that the difference is negligible. This gives us reason to assert that the analytical solution gives very good accuracy in the general case.

For greater clarity, we present the calculation results in the case of zero temperature, when $w_1=1$ and $w_2=w_3 =0$, using the results given in Appendix 2. We performed an analytical calculation according to \eqref{dma45}--\eqref{dma48} and numerical calculations of the longitudinal polarization $P_z(t)$ for the dimensionless phenomenological parameters $W=0.2$ and $\gamma=0.14$ to compare the results. We assume that the transmission rates are symmetric: $W_{23}=W_{32}=0.1$. Thus $\Gamma=(W-\gamma)/2=0.03$ and $\Gamma_0=(W+\gamma)/2=0.17$. The symmetric choice $W_{ik} = W_{ki} = W/2$ used below corresponds to the high-temperature limit of the detailed-balance relation $W_{mk}/W_{km} = \exp{((E_k - E_m)/T)}$, which is justified near the anticrossing, where $\omega_{23} \ll T$ for any realistic temperature.

\begin{figure}[htbp]
	\centering
	\includegraphics[width=\linewidth]{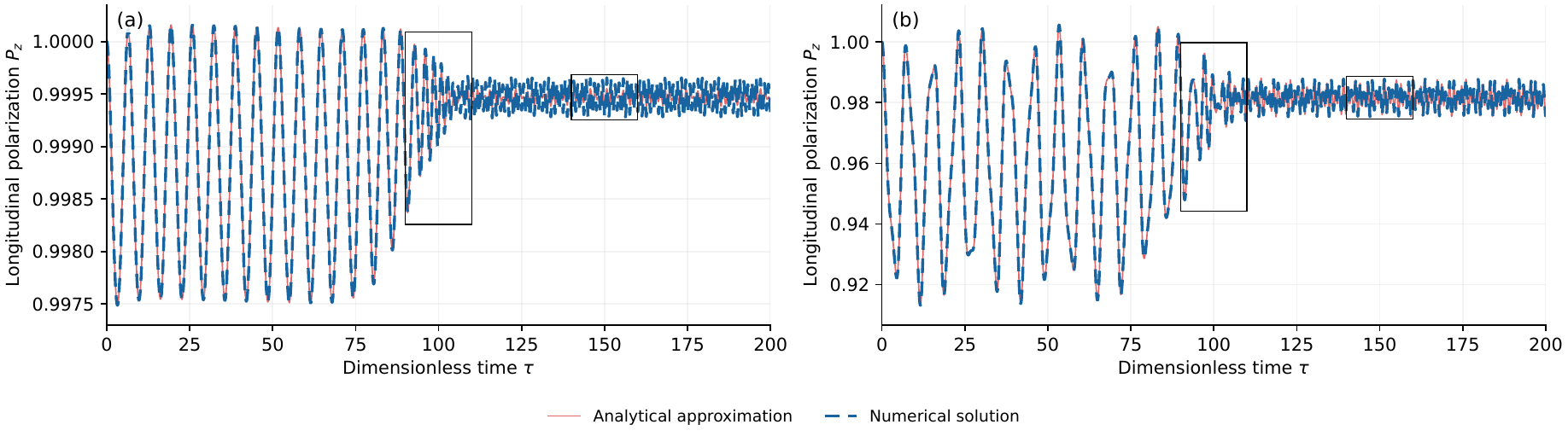}
	\caption{Longitudinal polarization for (a) $\theta=0.05\pi$ and (b) $\theta=\pi/8$. Thin red solid curves show the analytical approximation, and thicker blue dashed curves show numerical integration of the same reduced reference model. Black rectangles mark $90\leq\tau\leq110$ and $140\leq\tau\leq160$, enlarged in Figs.~\ref{fig3} and \ref{fig4}.}
	\label{fig2}
\end{figure}

Fig. \ref{fig2} shows a comparison of the results of analytical and numerical calculations for two angles $\theta = 0.05\pi$ and $\theta = 0.125\pi$, when the analytical expressions for energy levels have good accuracy\footnote{Note the different vertical scales of the two panels on this and following figures.}.
%%%%%%%%%%%%%%%%%%%%%%%%%%%%%%%%%%%%%%%%%%%%%%%%%

%%%%%%%%%%%%%%%%%%%%%%%%%%%%%%%%%%%%%%%%%%%%%%%%%%%%%%%%%%%%%%%
As can be seen, for external field values that do not reach the region of anticrossing of energy levels ($\tau < 75$), the analytical solutions and numerical calculations coincide for both angles. In the region of anticrossing of levels, the amplitude of oscillations decreases sharply and becomes very small for external fields located far from the anticrossing region. This significant diminishing in the amplitude of the oscillations is explained by the suppression of the solution for the density matrix \eqref{dma41} and \eqref{dma42} by the small parameters $\lambda_1$ and $\lambda_2$. At the same time, we can see that for a small angle $\theta = 0.05\pi$, the expansion parameters $\lambda_1$ and $\lambda_2)$ do not satisfy inequality \eqref{dma38} in the region of the level anticrossing.

\begin{figure}[htbp]
	\centering
	\includegraphics[width=\linewidth]{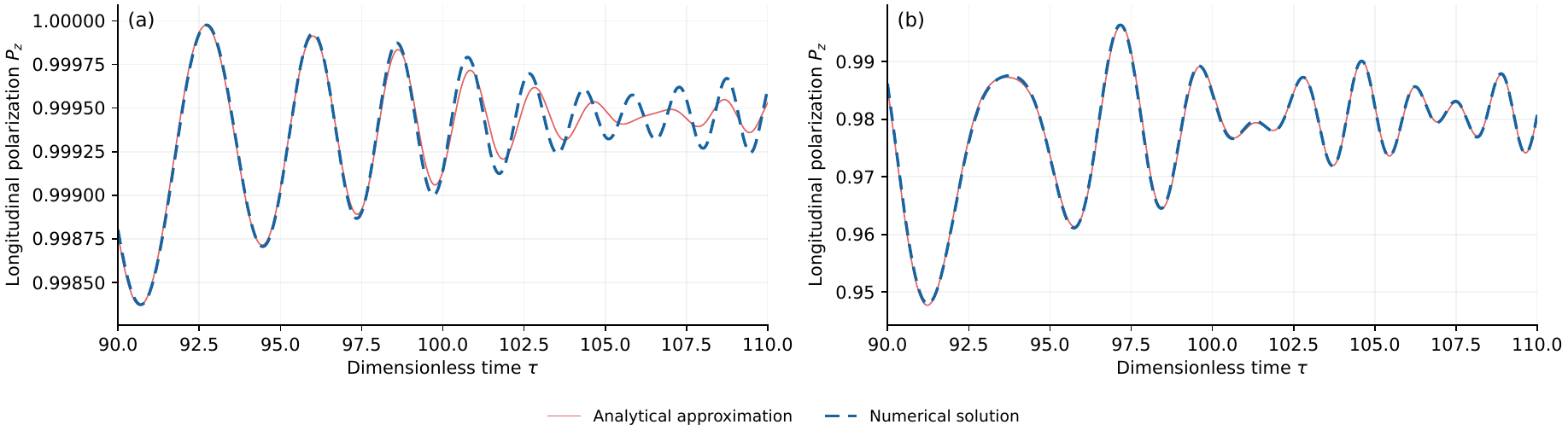}
	\caption{Central interval $90\leq\tau\leq110$ of the longitudinal polarization for (a) $\theta=0.05\pi$ and (b) $\theta=\pi/8$. The passage illustrates the loss of a uniformly small expansion parameter for the smaller angle.}
	\label{fig3}
\end{figure}

\begin{figure}[htbp]
	\centering
	\includegraphics[width=\linewidth]{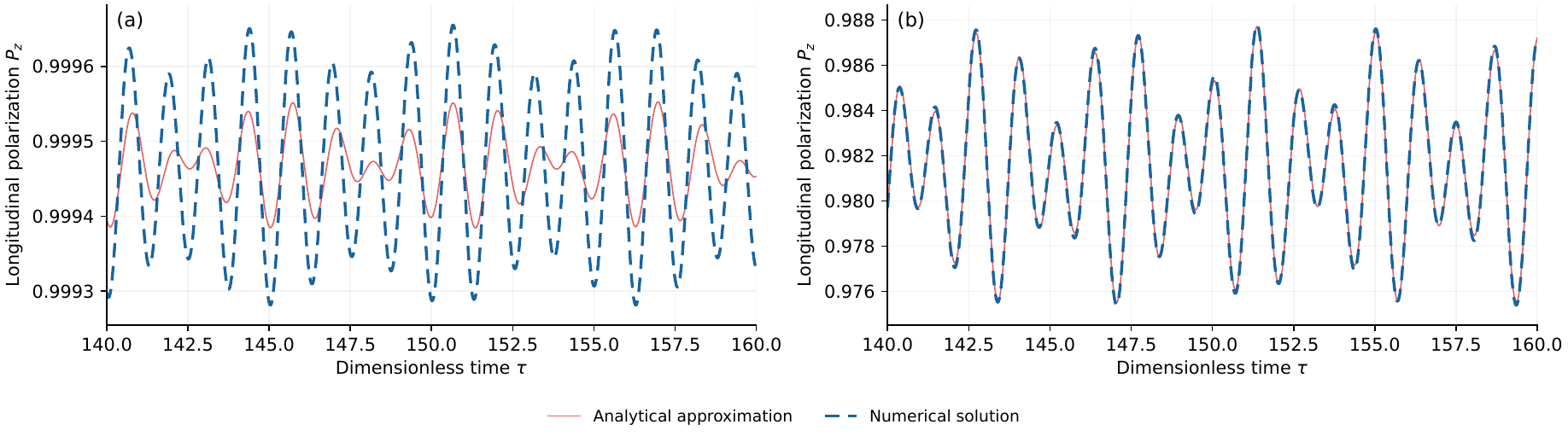}
	\caption{Late interval $140\leq\tau\leq160$ of the longitudinal polarization for (a) $\theta=0.05\pi$ and (b) $\theta=\pi/8$. The nearly saturated field gives common late-time frequency components, while the preceding passage can leave different amplitudes and phase offsets, especially in panel (a).}
	\label{fig4}
\end{figure}

Indeed, for $\theta=0.05\pi$ and $0.125\pi$, the parameters $(|\lambda_1|,\lambda_2)$ are approximately equal to $(0.441,0.0897)$ and $(0.0646,0.0355)$, respectively. Thus, the parameter $\lambda_1$ does not satisfy the smallness conditions for the angle $\theta=0.05\pi$. The divergence of analytical results and numerical calculation for the angle $\theta=0.05\pi$ and coincidence for  angle $\theta=0.125\pi$ are shown in Figs. \ref{fig3} and \ref{fig4}. It should be emphasized that the component $\rho_{11}(t)$ remains virtually unchanged, even if the adiabaticity condition for $\rho_{23}$ is violated, since the level separations $\omega_{12}$ and $\omega_{13}$ never become small.

%%%%%%%%%%%%%%%%%%%%%%%%%%%%%%%%%%%%%%%%%%%%%%%%%

%%%%%%%%%%%%%%%%%%%%%%%%%%%%%%%%%%%%%%%%%%%%%%%%%

In the late interval of Fig.~\ref{fig4}, the field is nearly constant and the basis-rotation couplings are small. The surviving polarization oscillations are then governed mainly by the $1$--$2$ and $1$--$3$ energy differences, which are common to the analytical and numerical descriptions used here. Accordingly, the late-time frequency components agree even for $\theta=0.05\pi$. The passage through the region of poor scale separation can nevertheless change the amplitudes and phase offsets of the coherences. Recovery of small local expansion parameters therefore restores agreement of the local oscillation frequencies but does not erase the earlier state error.

\section{Conclusion}

In this paper, we applied the previously introduced adiabatic representation to the formalism of the density matrix of non-stationary systems and derived the Liouville equation in this representation. The adiabatic representation is very convenient for describing quantum systems with a time-dependent Hamiltonian when it is impossible to apply the standard interaction representation. In this case, we need to define the density matrix in a non-stationary basis. The description of the density matrix in a non-stationary basis was also carried out in some articles (see, for example, \cite{Demirplak}), but the approach used there did not take into account the kinematic effects that lead to effective interaction. The adiabatic representation for the density matrix formalism is relevant, since only the density matrix allows us to correctly describe relaxation processes, which is very important when studying real open systems.

We illustrate the analytical construction using two identical, spatially separated spin-$1/2$ particles coupled by the magnetic dipole--dipole interaction. The selected polarization examples show how the separation between basis-rotation rates and level spacings affects the approximation within the reduced reference model. Connections to shortcuts to adiabaticity (see, e.g., \cite{C,Bernardo}) and systems with avoided crossings (see, e.g., \cite{Dodin}) motivate further study; the present examples do not establish general accuracy bounds.

Figures~\ref{fig2} and \ref{fig3} illustrate longitudinal polarization within the stated reduced model and relaxation assumptions. They emphasize the role of scale separation in the analytical approximation, rather than a numerical survey of driving protocols. Polarization alone is not a quantitative measure of entanglement. The general scope
of the construction follows from its derivation and assumptions, not from the number of examples displayed.

\appendix
\section{Series for the evolution operator}

Let us write down the series for the operator $U_R(t)$ in a form as
\begin{align}\label{a1}
  {\widehat U}_R=& 1 - \Gamma\int\limits_0^t\left(\cos 2\Phi_{ik}(t')\sigma_z +
   \sin 2\Phi_{ik}(t')\sigma_x\right)dt' + \\
   + &\Gamma^2\int\limits_0^t\int\limits_0^{t'}dt'dt''\left(\cos 2\Phi_{ik}(t')\sigma_z + \sin 2\Phi_{ik}(t')\sigma_x\right)\left(\cos 2\Phi_{ik}(t'')\sigma_z + \sin 2\Phi_{ik}(t'')\sigma_x\right) + O(\Gamma^3) \nonumber
\end{align}

Note that in $\Phi_{ik}(t)$ we have a rapidly oscillating integrand function $\cos\phi_{ik}(t)$, which makes it possible to make a good approximation:
\[
 \Phi_{ik}(t)=\int\limits_0^t{\dot \vartheta}(t')\cos\phi_{ik}(t')dt' \approx
 \frac{{\dot \vartheta}(t')}{\omega_{ik}(t')}\sin\phi_{ik}(t')\Bigr|_{0}^{t} =
 \frac{{\dot \vartheta}(t)}{\omega_{ik}(t)}\sin\phi_{ik}(t)
\]
since ${\dot \vartheta}(0) =0$, and we have taken into account that $d\phi_{ik}(t)/dt = \omega_{ik}(t)$. We can see that $|\varphi_{ik}(t)|\ll 1$. This allows us to decompose $\cos 2\Phi_{ik}(t)$ and $\sin 2\Phi_{ik}(t)$ to obtain approximate expressions for the corresponding integrals without taking into account the explicit form of the functions.
\[
\int\limits_0^t\cos 2\Phi_{ik}(t')dt' \approx \int\limits_0^t\left(1-\frac{1}{2}\left(\frac{2{\dot \vartheta}(t')}{\omega_{ik}(t')}\sin\phi_{ik}(t')\right)^2\right)dt' \approx
\]
\[
\approx t-\frac{1}{2}\left(\frac{2{\dot \vartheta}(t)}{\omega_{ik}(t)}\right)^2 \int\limits_0^t\sin^2\phi_{ik}(t') dt' =
  \left(1-\frac{{\dot \vartheta}^2(t)}{\omega^2_{ik}(t)}\right)t.
\]
Correspondingly
\[
 \int\limits_0^t\sin 2\Phi_{ik}(t')dt' \approx  \int\limits_0^t
 \frac{2{\dot \vartheta}(t')}{\omega_{ik}(t')}\sin\phi_{ik}(t')dt' \approx
 2\frac{{\dot \vartheta}(t)}{\omega^2_{ik}(t)}.
\]
In the second term of the decomposition, the integral function is transformed into the following expression:
\[
  \cos 2(\Phi_{ik}(t')-\Phi_{ik}(t'')) -i\sigma_y  \sin 2(\Phi_{ik}(t')-\Phi_{ik}(t'')).]
\]
This expression makes it possible to perform the following transformations:
\[
\!\!\! \Phi_{ik}(t')-\Phi_{ik}(t'')=2\int\limits_{t''}^{t'} {\dot \vartheta}(\xi)\cos\phi_{ik}(\xi)d\xi \approx 2{\dot \vartheta}(t')\left(\frac{\sin\phi_{ik}(t')}{\omega_{ik}(t')}- \frac{\sin\phi_{ik}(t'')}{\omega_{ik}(t'')}\right).
\]
Let's denote $t'-t''=\tau$, and considering that the frequencies of $\omega_{ik}(t)$ vary slowly over time, we can write
\[
 \!\!\! \sin\phi_{ik}(t'-\tau)\approx \sin\left(\phi_{ik}(t')-\frac{d\phi_{ik}}{dt'}\tau\right) \approx  \sin\phi_{ik}(t') - \cos\phi_{ik}(t') \omega_{ik}(t')\tau.
\]
Here we have assumed $\omega_{ik}(t')\tau \ll 1$. Keeping only the first order of the small parameter, we have
\[
 \Phi_{ik}(t')-\Phi_{ik}(t'-\tau) \approx 2 {\dot\vartheta}(t')\tau\cos\phi_{ik}(t').
\]
Now we can see that the smallness of the value of $\Phi_{ik}(t')-\Phi_{ik}(t'')$ is determined by the small parameter ${\dot\vartheta}(t')\tau$, which is valid up to $\omega_{ik}(t)\tau \sim 1$. Therefore, we can change $\tau \to \omega^{-1}_{ik}$ and decompose both the cosine and the sine down to the second order in a small parameter, and get
\[
\int\limits_0^t\int\limits_0^{t'}dt'dt''\cos 2(\Phi_{ik}(t')-\Phi_{ik}(t'')) \approx
\int\limits_0^t\int\limits_0^{t'}dt'd\tau \left(1-4 \left(\frac{{\dot\vartheta}(t')}{\omega_{ik}}\right)^2\cos^2\phi_{ik}(t')\right)=
\]
\[
=\frac{1}{2}t^2-4\int\limits_0^t\left(\frac{{\dot\vartheta}(t')}{\omega_{ik}(t')}\right)^2 \cos^2\phi_{ik}(t') t' dt'=
\]
\[
=\frac{1}{2}t^2-2\int\limits_0^t \left(\frac{{\dot\vartheta}(t')}{\omega_{ik}(t')}\right)^2(1+\cos2\phi_{ik}(t'))t' dt' \approx \frac{1}{2}t^2\left(1-\frac{2{\dot\vartheta}^2(t')}{\omega^2_{ik}(t')}\right).
\]
Since $\cos\phi_{ik}(t')$ is a rapidly oscillating function, and our decomposition assumes that $\omega_{ik}(t)\tau\sim 1$, we need to choose an upper limit of the integral equal to $t\approx\omega^{-1}_{ik}(t)$, and as a result we get
\[
\int\limits_0^t\int\limits_0^{t'}dt'dt''\cos 2(\Phi_{ik}(t')-\Phi_{ik}(t'')) \approx
\frac{1}{2}t^2 -\frac{1}{12}\frac{{\dot\vartheta}^2(t)}{\omega^4_{ik}(t)}.
\]
The second addition is
\[
\int\limits_0^t\int\limits_0^{t'}dt'dt''\sin 2(\Phi_{ik}(t')-\Phi_{ik}(t'')) \approx
2\int\limits_0^t{\dot\vartheta}(t')\cos\phi_{ik}(t'){t'}^2 dt'\approx \frac{2{\dot\vartheta}(t)}{\omega^3_{ik}(t)}.
\]
Finally, with an accuracy of the second order in a small parameter, we obtain the following expression for the desired operator
\begin{equation}\label{a2}
\!\!\!{\widehat U}_R =1 - \Gamma \left[\left(1-\frac{{\dot \vartheta}^2(t)}{\omega^2_{ik}(t)}\right)t\sigma_z +2\frac{{\dot \vartheta}(t)}{\omega^2_{ik}(t)} \sigma_x\right] + \frac{1}{2}\Gamma^2\left[t^2\left(1-\frac{{\dot \vartheta}^2(t)}{\omega^2_{ik}(t)}\right)^2 -i\frac{2{\dot\vartheta}(t)}{\omega^3_{ik}(t)}\sigma_y \right] .
\end{equation}
Here, we have taken into account that with an accuracy of the second order in a small parameter is valid $1 + 2\lambda \approx (1 + \lambda)^2$.

\section{Thermal equilibrium and laboratory preparation}

For the interacting Hamiltonian, the thermal state at the initial instant is
\begin{equation*}
	\rho_{\mathrm{th}}(0)=
	\frac{e^{-H(0)/T}}{\Tr e^{-H(0)/T}},
	\qquad
	\widetilde\rho_{\mathrm{th}}(0)=
	\Theta^\dagger(0)\rho_{\mathrm{th}}(0)\Theta(0).
\end{equation*}
It is diagonal in an exact energy eigenbasis, not generally in the laboratory basis $\{|q_k\rangle\}$. At zero temperature it approaches the normalized projector onto the ground eigenspace, rather than necessarily onto $|q_1\rangle$. The examples in this paper instead use the laboratory preparation $\rho(0)=|q_1\rangle\langle q_1|$, independently of a thermal equilibrium assumption. Its transformed density matrix is
\begin{equation*}
    \rho_{kk}(0)= \frac{1}{Z(T)}e^{-\varepsilon_k/T}, \quad Z(T)=\sum_k e^{-\varepsilon_k/T}=
    e^{-\varepsilon_1/T}\left(1+e^{\omega_{12}/T}+e^{\omega_{13}/T}\right).
\end{equation*}
The matrix elements of the transformed density matrix in this case can be written as follows:
\begin{eqnarray*}
  {\tilde \rho}_{11} &=& F(T)\left[1+e^{\omega_{12}/T}+ \cos2\vartheta_1\left(1-e^{\omega_{12}/T}\right)\right], \\
  {\tilde \rho}_{22} &=&\frac{1}{2}F(T)\left[\left(1+e^{\omega_{12}/T}- \cos2\vartheta_1\left(1-e^{\omega_{12}/T}\right)\right)(1+\cos2\vartheta_2)+ 2(1-\cos2\vartheta_2)e^{\omega_{13}/T}\right], \\
  {\tilde \rho}_{33} &=&\frac{1}{2}F(T)\left[\left(1+e^{\omega_{12}/T}- \cos2\vartheta_1\left(1-e^{\omega_{12}/T}\right)\right)(1-\cos2\vartheta_2)+ 2(1+\cos2\vartheta_2)e^{\omega_{13}/T}\right], \\
  {\tilde \rho}_{12} &=& {\tilde \rho}_{21} = -F(T)\sin2\vartheta_1\cos\vartheta_2 \left(1-e^{\omega_{12}/T}\right),\\
  {\tilde \rho}_{13} &=& {\tilde \rho}_{31} = F(T)\sin2\vartheta_1\sin\vartheta_2 \left(1-e^{\omega_{12}/T}\right), \\
  {\tilde \rho}_{23} &=& {\tilde \rho}_{32} =-\frac{1}{2}F(T)\left[1+e^{\omega_{12}/T}- \cos2\vartheta_1\left(1-e^{\omega_{12}/T}\right)-2e^{\omega_{13}/T}\right],
\end{eqnarray*}
where
\[
  F^{-1}(T)=2\left(1+e^{\omega_{12}/T}+e^{\omega_{13}/T}\right).
\]
In the case of zero temperature $T=0$ we have $w_1=1$ and $w_2=w_3=0$, and the transformed density matrix takes the form
\begin{equation}\label{dma_a2}
\!\!\!  \tilde{\rho}_{+1}(0)\!=
       \frac{1}{2} \left(
  \begin{array}{ccc}
 \!\!\!\!1+ \cos2\vartheta_{0,1} &\!\! -\sin 2\vartheta_{0,1}\cos\vartheta_{0,2} &\!\!\! \sin 2\vartheta_{0,1}\sin\vartheta_{0,2}\!\!\!\\
 \!\!\!\!-\sin 2\vartheta_{0,1}\cos\vartheta_{0,2} &\!\!\frac{1}{2}(1-\cos 2\vartheta_{0,1})
 (1+\cos 2\vartheta_{0,2})&\!\!\!-\frac{1}{2}(1-\cos 2\vartheta_{0,1})\sin 2\vartheta_{0,2} \!\!\! \\
 \!\!\!\!\sin 2\vartheta_{0,1}\sin\vartheta_{0,2} &\!\!-\frac{1}{2}(1-\cos 2\vartheta_{0,1})\sin 2\vartheta_{0,2}
 &\!\!\! \frac{1}{2}(1-\cos 2\vartheta_{0,1})(1-\cos 2\vartheta_{0,2}) \!\!\!\\
  \end{array}
  \right).
\end{equation}

\section{Initial transformed density matrix}

In the following it is useful to express the components of the transformed density matrix at the initial time in terms of the transformation parameters \eqref{dma14} and \eqref{dma15}.
\begin{align*}
	\widetilde\rho_{11}(0)
	&=\frac12\left[w_1+w_2+(w_1-w_2)\cos(2\vartheta_{0,1})\right],\\
	\widetilde\rho_{22}(0)+\widetilde\rho_{33}(0)
	&=w_3+\frac12\left[w_1+w_2-(w_1-w_2)\cos(2\vartheta_{0,1})\right],\\
	\widetilde\rho_{22}(0)-\widetilde\rho_{33}(0)
	&=\frac12\left[w_1+w_2-(w_1-w_2)\cos(2\vartheta_{0,1})-2w_3\right]
	\cos(2\vartheta_{0,2}),\\
	\widetilde\rho^s_{12}(0)
	&=-\frac12(w_1-w_2)\sin(2\vartheta_{0,1})\cos\vartheta_{0,2},\\
	\widetilde\rho^s_{13}(0)
	&=\frac12(w_1-w_2)\sin(2\vartheta_{0,1})\sin\vartheta_{0,2},\\
	\widetilde\rho^s_{23}(0)
	&=\frac14\left[2w_3-w_1-w_2+(w_1-w_2)\cos(2\vartheta_{0,1})\right]
	\sin(2\vartheta_{0,2}).
\end{align*}
Here the argument $0$ denotes the initial instant; for a shifted
observation window it is replaced by $\tau_i$.

\end{document}